\documentclass[twocolumn]{aastex631}

\usepackage{hyperref}
\usepackage{cancel}
\usepackage{subfigure}
\shorttitle{NGC346 internal kinematics}
\shortauthors{Sabbi et al.}
\graphicspath{{./}{figures/}}

\begin{document}

\title{The internal proper motion kinematics of NGC~346: past formation and future evolution}

\author[0000-0003-2954-7643]{E. Sabbi}
\affiliation{Space Telescope Science Institute, 3700 San Martin Dr, Baltimore, MD 21218, USA}

\author[0000-0002-6091-7924]{P. Zeidler}
\affil{AURA for the European Space Agency (ESA), ESA Office, Space Telescope Science Institute, 3700 San Martin Drive, Baltimore, MD 21218, USA}

\author[0000-0001-7827-7825]{R.P. van der Marel}
\affil{Space Telescope Science Institute, 3700 San Martin Dr, Baltimore, MD 21218, USA}

\author{A. Nota}
\affil{European Space Agency (ESA), ESA Office, Space Telescope Science Institute, 3700 San Martin Drive, Baltimore, MD 21218, USA}

\author[0000-0003-2861-3995]{J. Anderson}
\affil{Space Telescope Science Institute, 3700 San Martin Dr, Baltimore, MD 21218, USA}

\author[0000-0001-8608-0408]{J.S. Gallagher}
\affil{Department of Astronomy, University of Wisconsin–Madison, 5534 Sterling, 475 North Charter Street, Madison, WI 53706, USA}

\author{D.J. Lennon}
\affil{Instituto de Astrofísica de Canarias,E-38 200 La Laguna, Tenerife, Spain}
\affil{Dpto. Astrofísica, Universidad de La Laguna, E-38 205 La Laguna, Tenerife, Spain}

\author[0000-0002-0806-168X]{L.J. Smith}
\affil{Space Telescope Science Institute, 3700 San Martin Dr, Baltimore, MD 21218, USA}

\author[0000-0002-5581-2896]{M. Gennaro}
\affil{Space Telescope Science Institute, 3700 San Martin Dr, Baltimore, MD 21218, USA}



\begin{abstract}
We investigate the internal kinematics of the young star-forming region NGC 346 in the Small Magellanic Cloud. We used two epochs of deep F555W and F814W Hubble Space Telescope ACS observations with an 11-year baseline to determine proper motions, and study the kinematics of different populations, as identified by their color-magnitude diagram and spatial distribution characteristics. The proper motion field of the young stars shows a complex structure with spatially coherent patterns. NGC 346 upper-main sequence and pre-main sequence stars follow very similar motion patterns, with the outer parts of the cluster being characterized both by outflows and inflows. The proper motion field in the inner $\sim 10$ pc shows a combination of rotation and inflow, indicative of inspiraling motion. The rotation velocity in this regions peaks at $\sim 3$ km/s, whereas the inflow velocity peaks at $\sim 1$ km/s. Sub-clusters and massive young stellar objects in NGC 346 are found at the interface of significant changes in the coherence of the proper motion field. This suggests that turbulence is the main star formation driver in this region. The similar kinematics observed in the metal-poor NGC~346 and the Milky Way star-forming regions suggest that the differences in the cooling conditions due to the different amounts of metallicity and dust density between the SMC and our Galaxy are too small to alter significantly the process of star clusters assembly and growth. The main characteristics of our findings are consistent with various proposed star cluster formation models.
\end{abstract}



\section{Introduction}
\label{sec:intro}

Regions of massive star formation (SF) are important to study for the insight they provide into the origin of the stellar mass distribution, the formation of massive stars ($M>8\, M_\odot$), and the formation and evolution of associations and star clusters. Nevertheless, the process whereby stars form in giant molecular clouds is still a poorly constrained problem.

The first few million years, during which stars are still forming and the region still contains a significant amount of gas, set the conditions for the subsequent evolution of the stellar system. This initial stage is regulated by gas and stellar dynamics, stellar evolution, and radiative transfer. However, the complex interplay between these quantities is poorly understood by theory and lacks sufficient observational constraints \citep[i.e.][]{Elmegreen2007, Price2009, Portegies2010, Krumholz2014, Kruijssen2019}.
\begin{figure*}
\plotone{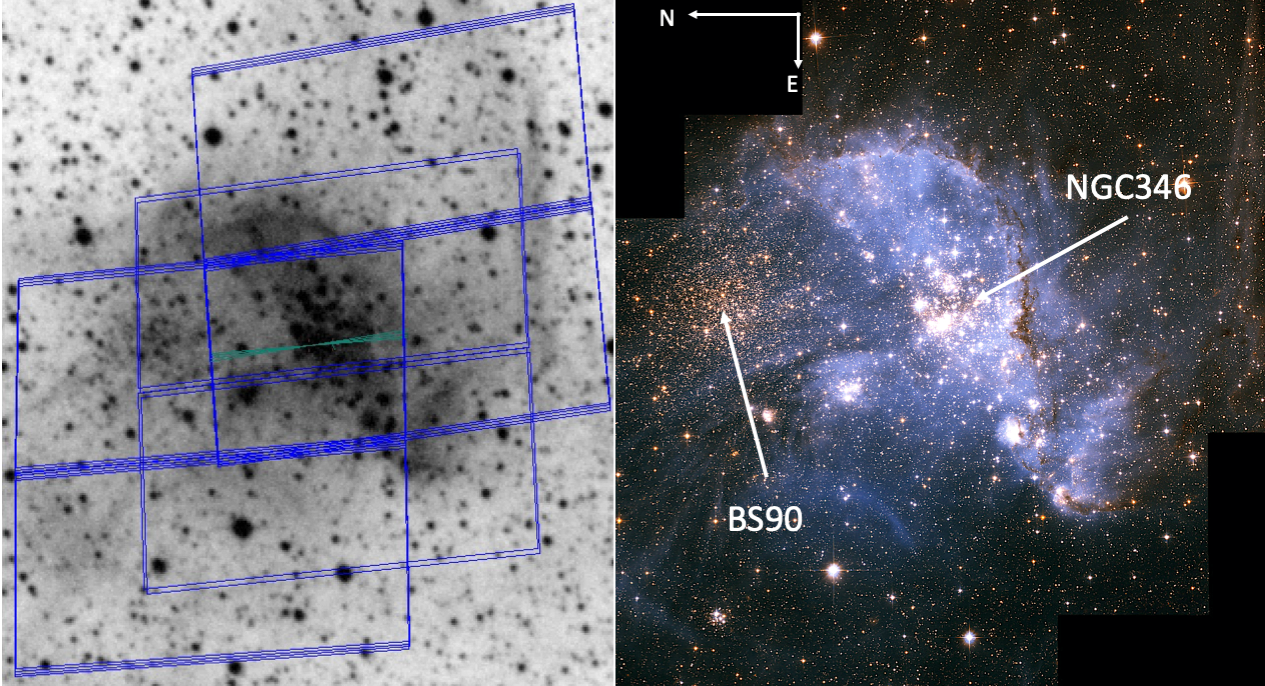}
\caption{{\it Left Panel:} Mosaic footprint of the area mapped by the programs GO-10248 and GO-13680 superimposed on the DSS image. {\it Right Panel:} HST color-composite image obtained combining the exposures taken by the program GO-10248. F555W is in blue and F814W in red. The positions of NGC~346 and BS90 are highlighted for convenience. North is left, East is down.}
\label{f:mosaic}
\end{figure*}

This shortfall of information leaves many essential and critical properties uncertain, such as the duration and the efficiency of the SF process, and hence the probability that a cluster will survive the rapid ejection of gas caused by its first supernova explosions \citep[][]{Carpenter2000, Lada2003, Chevance2020}.
The internal kinematics of young star clusters and associations carries the signature of the process that led to the systems' formation. Radial velocity and proper motion (PM) studies can gather information on the dynamical state of gas and stars, and test different models of SF  \citep{Elmegreen2002, Parker2014}. Currently, theories describing the onset and development of SF can be divided into two competing scenarios: a rapid process, which proceeds on a dynamical timescale, comparable to the freefall collapse of the molecular cloud \citep[e.g.,][]{Elmegreen2000, Dobbs2011, Hartmann2012, Grudic2018, Jeffreson2018}; and a slow one that persists for multiple freefall timescales \citep[e.g. ][]{McKee1989, Tan2006, Krumholz2020}

Astrometric measurements by ESA’s Gaia spacecraft \citep{Gaia2016} have already provided spatial and kinematic clustering properties for several star-forming regions in the Milky Way, including Orion \citep{Grossschedl2018, Kounkel2018, Getman2019, Kuhn2019}, Taurus \citep{Luhman2018, Galli2019}, $\rho$ Oph \citep{Canovas2019}, Serpens \citep{Herczeg2019}, NGC~6530 \citep{Kuhn2019}, and IC 5070 \citep{Kuhn2020}.

In this work we extend this type of investigation for the first time to a low metallicity environment \citep[$Z=0.2 Z_\odot$, ][]{Russell1992, Rolleston2003, Hunter2007}, such as the Small Magellanic Cloud (SMC), to explore whether different global conditions (e.g., as large-scale dynamics and metallicity) can affect the duration of SF.

As the brightest and largest SMC star-forming region, NGC~346, located in the northern part of the galaxy bar, is one of the best-studied extra-galactic young clusters. In an attempt to better understand the formation mechanism and early evolution of NGC~346, here we take advantage of the {\it Hubble Space Telescope}'s (HST) exquisite astrometric capabilities and longevity to measure the PM displacements of high, intermediate, and low-mass ($\sim 8$ to $\sim 1\, {\rm M_\odot}$) stars and infer the system's internal kinematics.

\begin{table*}[htb]
    \centering
    \begin{tabular}{cccccccccc}
    \hline
    GO & Date & Filter & Pointing & N $\times$ Exp. time (s) & GO & Date & Filter & Pointing & N $\times$ Exp. time (s) \\
    \hline
    10248 & July 2004 & F555W & Center & 1 $\times$ 380 & 13680 & July 2015 & F555W & Center & 3 $\times$ 450 \\
          &           &       & North  & 4 $\times$ 456 &       &           &       & North  & 4 $\times$ 450 \\
          &           &       & South  & 4 $\times$ 483 &       &           &       & South  & 4 $\times$ 450 \\
          &           & F814W & Center & 1 $\times$ 380 &       &           & F814W & Center & 2 $\times$ 450 \\
          &           &       & North  & 4 $\times$ 484 &       &           &       & North  & 4 $\times$ 450 \\
          &           &       & South  & 4 $\times$ 450 &       &           &       & South  & 4 $\times$ 450 \\
    \hline
    \end{tabular}
    \caption{List of the observations analyzed in this paper. The first and sixth columns report the proposal IDs, the observing dates are listed in columns two and seven, the used filters are in columns three and eight, the covered regions in columns four and nine. Exposure times are in columns five and ten.}
    \label{t:obs}
\end{table*}

The left panel of Fig.~\ref{f:mosaic} shows the footprint of the HST mosaic superimposed on the Digitized Sky Survey (DSS) image, while the right panel shows the HST color-composite image. Here NGC~346 appears as a vast agglomeration of bright white stars, still embedded in diffuse ionized gas (H$\alpha$ and N{\sc ii}, shown in blue), extending over $\sim 50$ pc from the northwest to the southeast. The star-forming region contains a ``plethora'' of massive stars \citep{Niemela1986, Massey1989, Evans2006, Dufton2019} that are ionizing the relatively isolated, large H{\sc ii} region N66. O and B stars are embedded in an extended halo of pre-main sequence (PMS) stars \citep[][]{Nota2006}, organized in several sub-clusters, clumps, and asterisms \citep{Sabbi2007, Gouliermis2014}.

In \citet{Cignoni2011} we studied the SF history of the region and found that SF started, with remarkable synchronization, about $\sim 6$ Myr ago. Then, progressing inward, it peaked $\sim 3$ Myr ago, and is now continuing at a lower rate. Similar conclusions were recently reached by \citet{Dufton2019}.

Extended CO clouds are still associated with NGC~346 \citep{Rubio2000, Muller2015, Neelamkodan2021}, and Spitzer observations revealed a multitude of young stellar objects \citep[YSOs, ][]{Simon2007, Sewilo2013} associated with the sub-clusters and asterisms identified by \citet{Sabbi2007}. \citet{Rubio2018} analyzed $HK$ spectra of three YSOs and concluded that these are likely $\sim 25-26\, M_\odot$ Class~{\sc I} type,  suggesting that, at least in the central cluster, star formation is still ongoing.

The bright clump of red stars to the north of NGC~346 is the core of the star cluster BS90. NGC~346 and BS90 are considered non-interacting, and their proximity is likely a simple visual alignment \citep{Bica1995}. Using the synthetic color-magnitude diagram (CMD) method initially introduced by \citet{Tosi1991}, \citet{Sabbi2007} concluded that the cluster is $4.5\pm 0.1\, {\rm Gyr}$ old, with metallicity $Z=0.02$, $E(B-V)=0.08$, distance modulus $(m-M)_0=18.9$, and a total mass $\sim 8 \times 10^4\,  {\rm M_\odot}$. The halo of BS90 partially overlaps in projection with NGC346, and  the two clusters are embedded in the SMC field, making it impossible to separate the different stellar populations using only their coordinates, or the characteristics of their CMDs.

\begin{figure}[htb]
\plotone{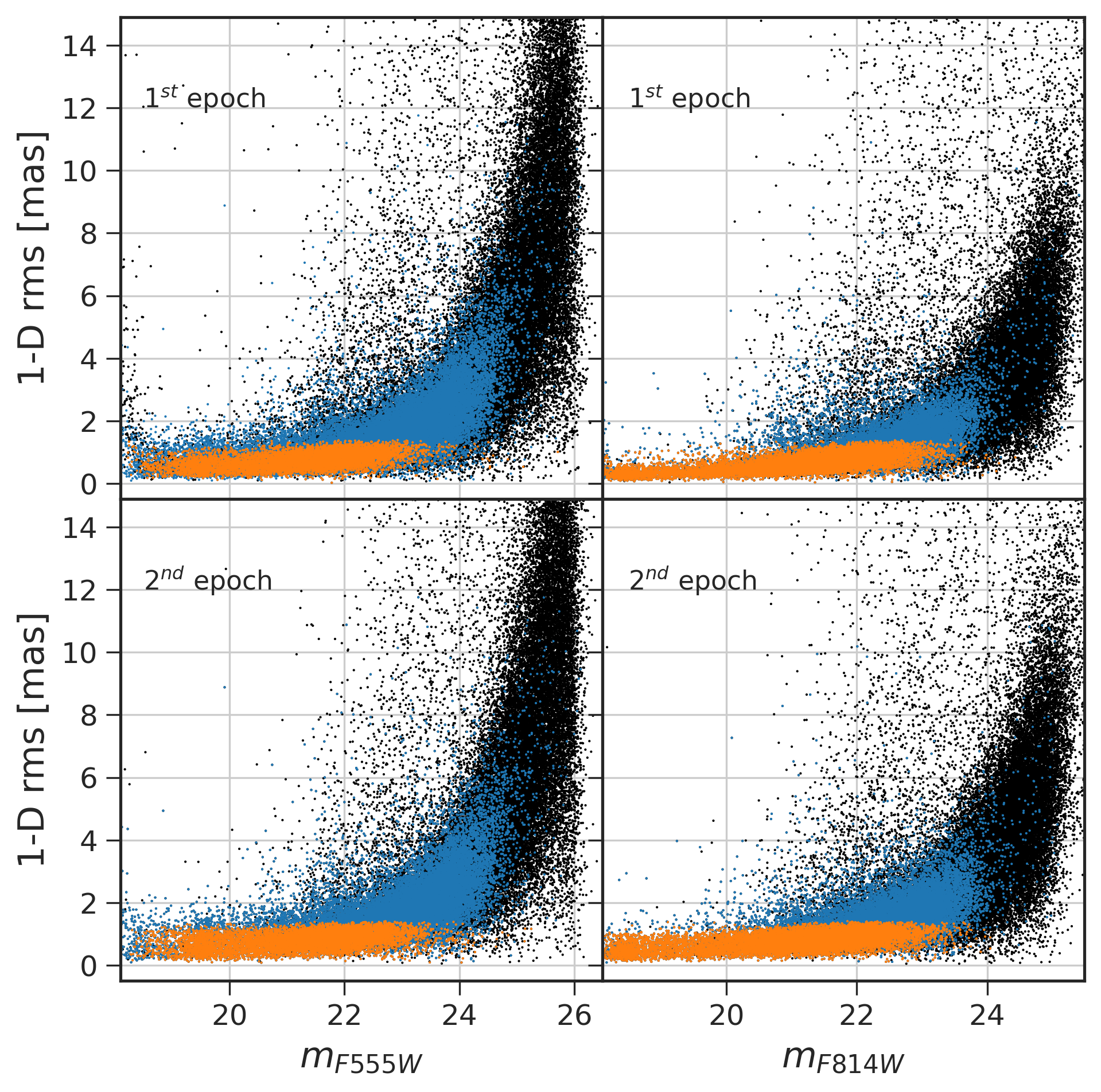}
\caption{1-D position rms in mas as a function of magnitude. The two {\it upper panels} show the results for the $1^{st}$ epoch, while the two {\it lower panels} show the results for the second epoch. The F555W filter is to the left and F814W to the right. In both filters and epochs, we marked all the sources with positional errors $<1$ mas and photometric errors $< 0.1$ in orange. Sources with photometric rms $<0.2$ in both filters and with positional rms in F814W $<3$ mas in both epochs are in blue.}
\label{f:errors}
\end{figure}

\begin{figure}[htb]
\plotone{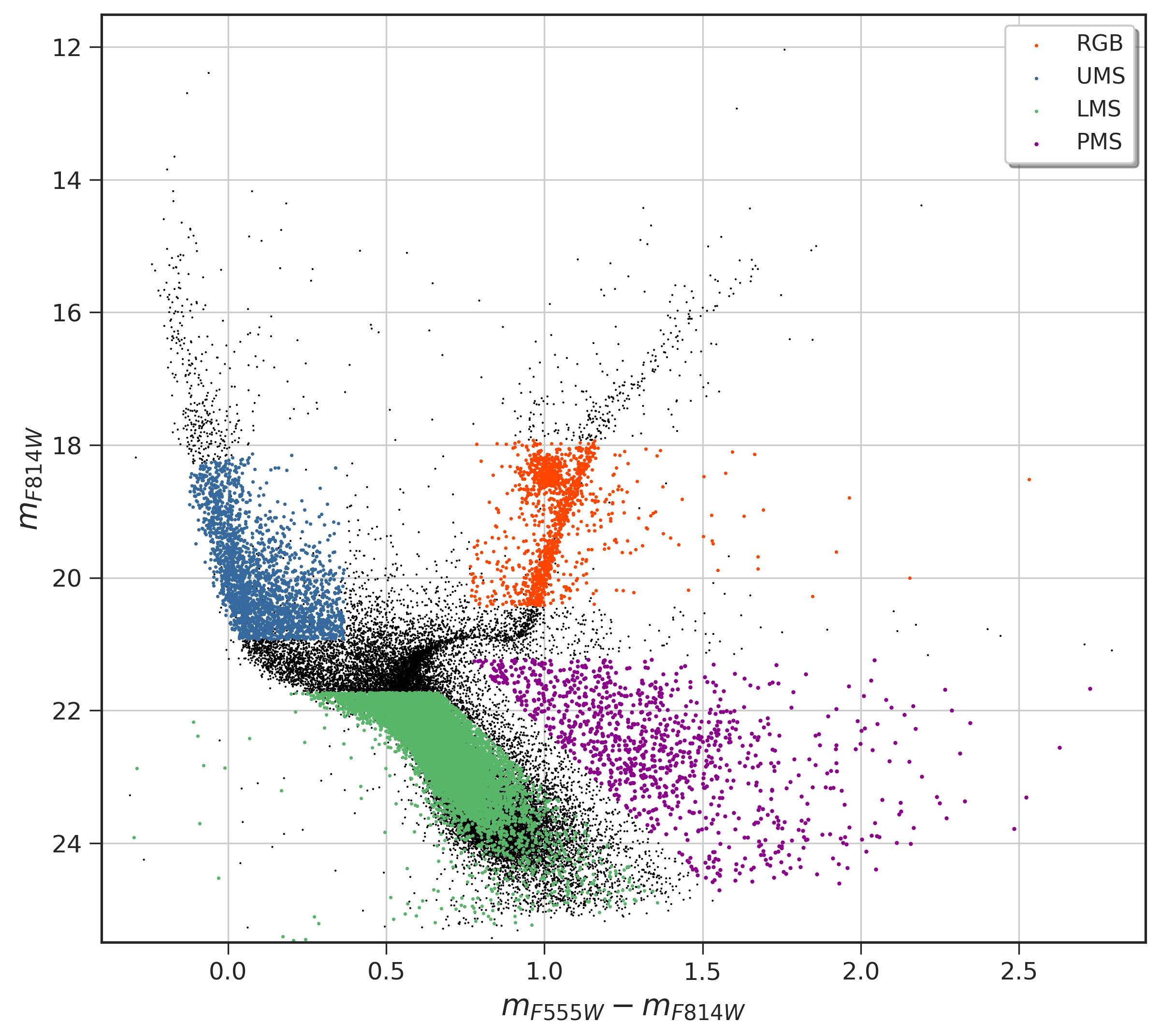}
\caption{The $m_{F555W}-m_{F814W}, m_{F814W}$ CMD of the stars detected in the NGC~346 region, with photometric rms $< 0.1$ in both F814W and F555W filters. Sources with positional rms $<0.1$ mas/yr in both X and Y directions and in both epochs have been highlighted as follow: RGB stars in the magnitude range $17.9<m_{F814W}<20.4$ and color $m_{F555W}-m_{F814W}>0.5$ are shown in red, and UMS stars in the magnitude ranges $17.9<m_{F814W}<20.9$ and color $m_{F555W}-m_{F814W}<0.1$ are shown in blue. LMS stars fainter than $m_{F814W}<21.9$ and to the left of the line connecting $m_{F555W}-m_{F814W}=0.7$, \,$m_{F814W}<21.9$ to $m_{F555W}-m_{F814W}=1.3$,\, $m_{F814W}<24.0$ are marked in green. PMS stars fainter than $m_{F814W}>21.2$, to the right of the line connecting $m_{F555W}-m_{F814W}=0.75$,\, $m_{F814W}<21.2$ to $m_{F555W}-m_{F814W}=1.5$,\, $m_{F814W}<24.0$, with photometric rms $<0.2$ in both F814W and F555W, and positional rms $<0.3$ mas/yr in both X and Y are shown in purple.}
\label{f:cmd}
\end{figure}

The paper is organized as follows: section~\ref{s:Obs} contains a description of the observations and the photometric analysis. We discuss the characteristics of the stellar populations found in the region in Section~\ref{stellar pops}. We present PM measurements in Section~\ref{s:PMs}, and in Section~\ref{s:kinem} we analyze the kinematics of the various populations. In Section~\ref{s:n346} we focus our attention on NGC~346 internal kinematics, and We present our conclusions in Section~\ref{s:Concl}.

\section{Observations and Data Reduction}
\label{s:Obs}

\begin{figure}[htb]
\plotone{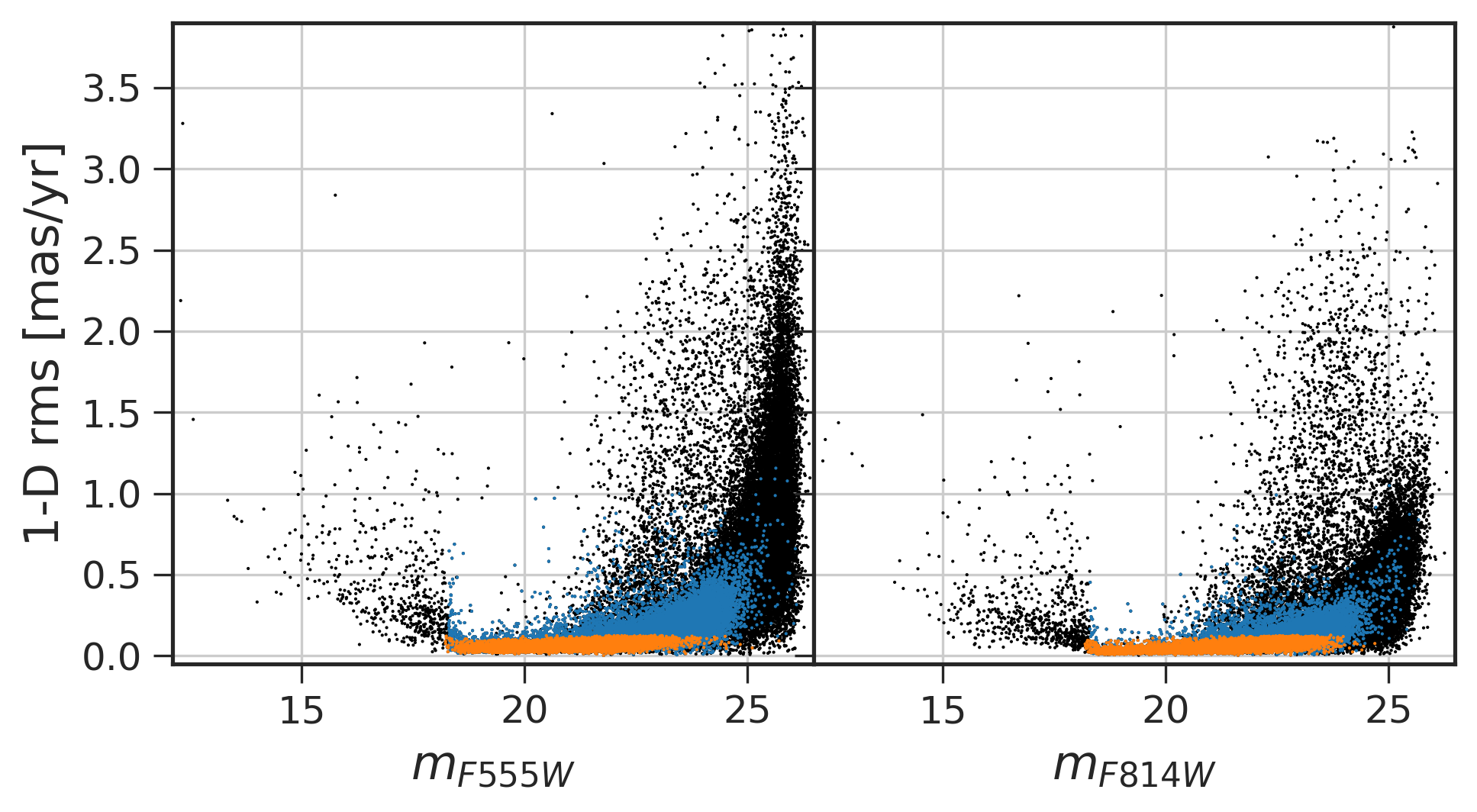}
\caption{1-D PM rms in mas/yr as a function of magnitude. The F555W filter is to the left and F814W to the right. The sources with positional errors $< 1$ mas and photometric errors $< 0.1$ are marked in orange. Sources with photometric rms $<0.2$ in both filters and and with positional rms in F814W $<3$ mas in both epochs are in blue.}
\label{f:PMerrors}
\end{figure}

\begin{figure*}[htb]
\plotone{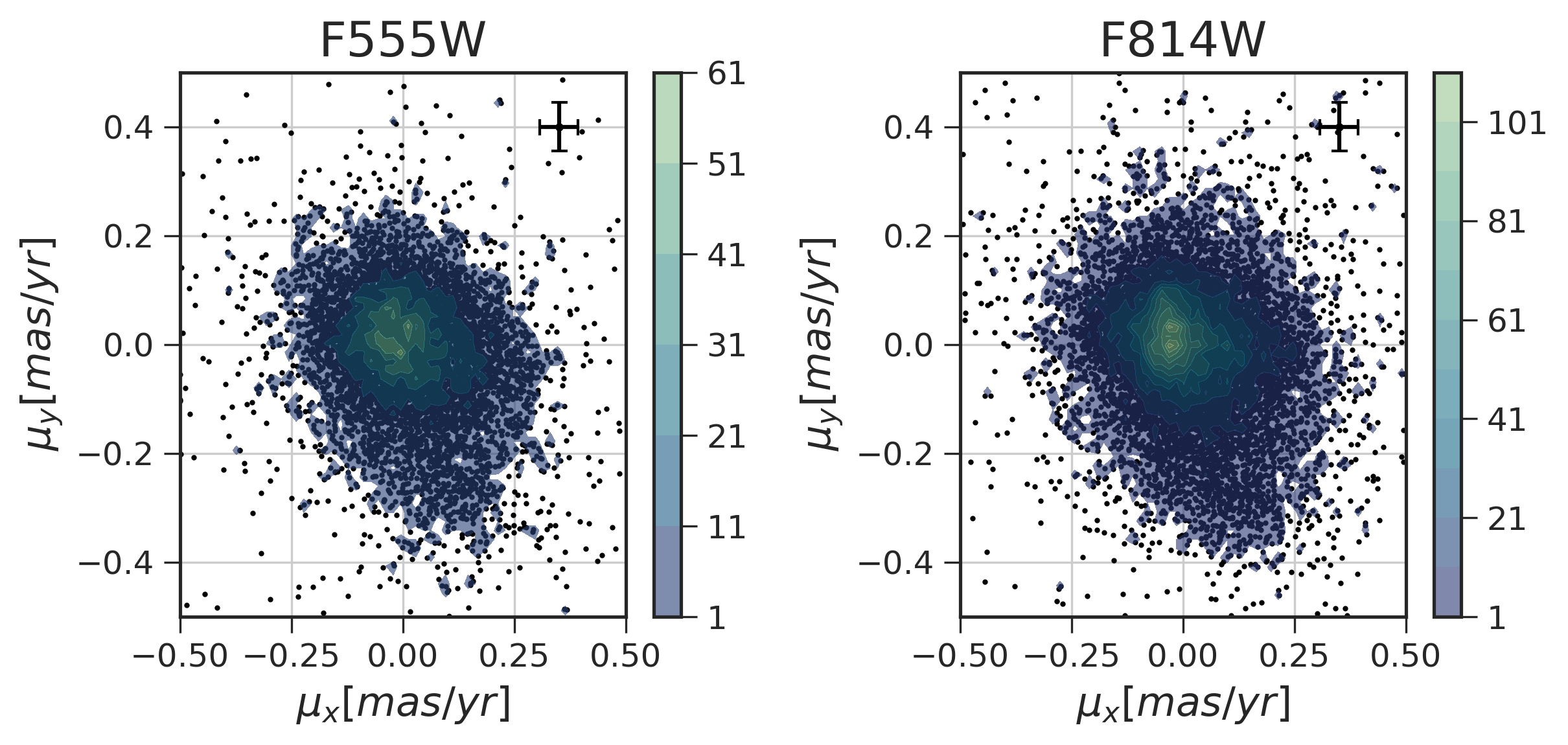}
\caption{The VPDs of the PMs measured in the NGC~346--N66 region. Each point corresponds to the proper motion of one star. Only stars that meet the astrometry and photometry selection criteria are considered. Measurement using the F555W filter are shown  in the {\it left panel}. The corresponding results for the F814W filter are shown in the {\it right panel}. Median errors along the X and Y directions are shown in the upper right corner.}
\label{f:VPD_tot}
\end{figure*}

(HST)'s Wide-Field Channel (WFC) of the Advanced Camera for Surveys (ACS; pixel scale $\sim 50$ mas pixel$^{-1}$) initially observed NGC~346 through the filters F555W and F814W in July 2004 (GO-10248, PI A. Nota). To measure the star-forming region's internal kinematics, we repeated these observations in July 2015, using the same orientation and filter set (GO-13680, PI. E. Sabbi). In both epochs, images were taken at three different pointing positions, each covering an area of $200\arcsec \times 200\arcsec$. In the first epoch, for each filter, a single long exposure was taken at the nominal center of the cluster. The other two pointings were spaced by $\sim 1\arcmin$ off the cluster's center towards the northeast and the southwest, respectively. A four-step dither pattern was applied in each filter for both the northern and the southern pointings to remove hot pixels, better sample the point spread function (PSF), and fill in the detector gap.
In the second epoch, the central pointing consisted of three-dithered exposures in F555W and two-dithered exposures in F814W.  As in the first epoch, we acquired four exposures per filter for both the northern and southern pointing. However, no dither stepping was applied to F555W filter observations covering the southern portion of the mosaic. Table~\ref{t:obs} summarizes the list of the observations. The observations analyzed in this paper can be downloaded from the Mikulski Archive for Space Telescopes (MAST) via  \dataset[10.17909/y1cf-m826]{https://doi.org/10.17909/y1cf-m826.}

We carried out the analysis of the data directly on the bias-subtracted, flat-field and charge-transfer-efficiency (CTE) corrected {\tt \_flc} exposures produced by the standard calibration pipeline CALACS v10.2.4. Compared to drizzled ({\tt \_drc}) images, {\tt \_flc} data have the advantage of not being re-sampled, thus providing a more direct representation of the astronomical scene. However, {\tt \_flc} exposures are still affected by geometric distortion. To take into account this effect, we created a  distortion-free reference frame using the geometric distortion correction for the ACS/WFC detectors described in \citet{Anderson2006}\footnote{The geometric distortion correction used in this paper is available for download at \href{url}{https://www.stsci.edu/~jayander/STDGDCs/}}, and we related the photometry and astrometry of each exposure to that frame.

We analyzed all images using the Fortran routine {\tt hst1pass}\footnote{\href{url}{https://www.stsci.edu/~jayander/HST1PASS/}} (Anderson et al. 2022, in prep.). The program performs a single pass of finding and measures each star in each exposure by fitting a library of spatially variable empirical PSFs\footnote{\href{url}{https://www.stsci.edu/~jayander/STDPSFs/}}, ignoring any contribution from neighbors.

Seasonal and orbital thermal changes can cause variations in the optical path length of {\it HST} up to a few microns within the timescale of an orbit \citep{Bely1993, Lallo2006}. These changes affect the telescope's focus and translate to small but measurable differences in the PSF from one exposure to another, possibly affecting the precision with which source positions are determined. To take into account these effects, in each image, we measured the average residuals after the subtraction of the empirical PSF-library from the bright (S/N $>$ 150), isolated (out to a 10-pixel radius), and non-saturated stars. We then applied, for each image, the correction to the PSF that minimizes its residuals and ran the photometry on all the sources with S/N $>$ 3 and without companions within 3 pixels. This approach allowed us to create for each exposure a catalog of X and Y coordinates in pixels, magnitudes, quality of the fit \citep[{\sc qfit}, as defined in ][]{Anderson2008} and $\chi^2$ of the fit.

The three pointing positions allowed us to observe the same stars both close and far from the readout amplifiers, thus verifying that in both epochs the CTE residuals were negligible both in stellar magnitudes and positions. We derived the photometric zero points by matching our observations to the NGC~346 photometric catalog published by \citet{Sabbi2007}, and available for download from VizieR\footnote{\href{url}{https://vizier.u-strasbg.fr/viz-bin/VizieR}}

We used a six-parameter linear transformation based on the cross-identified, well-measured, and non-saturated stars to match all the exposures in a common photometric and astrometric reference frame for each epoch$+$filter combination. We obtained the initial reference frame positions by averaging the single-image positions. For both filters, the median 1-D rms in the position of the bright stars ($18<m<21$) is less than 0.5 mas (=0.01 pxl) in the first epoch and 0.7 mas (0.015 pxl) in the second epoch. Fig.~\ref{f:errors} shows the position rms as a function of magnitude.

\section{Stellar Content}
\label{stellar pops}

\begin{figure*}[htb]
\plotone{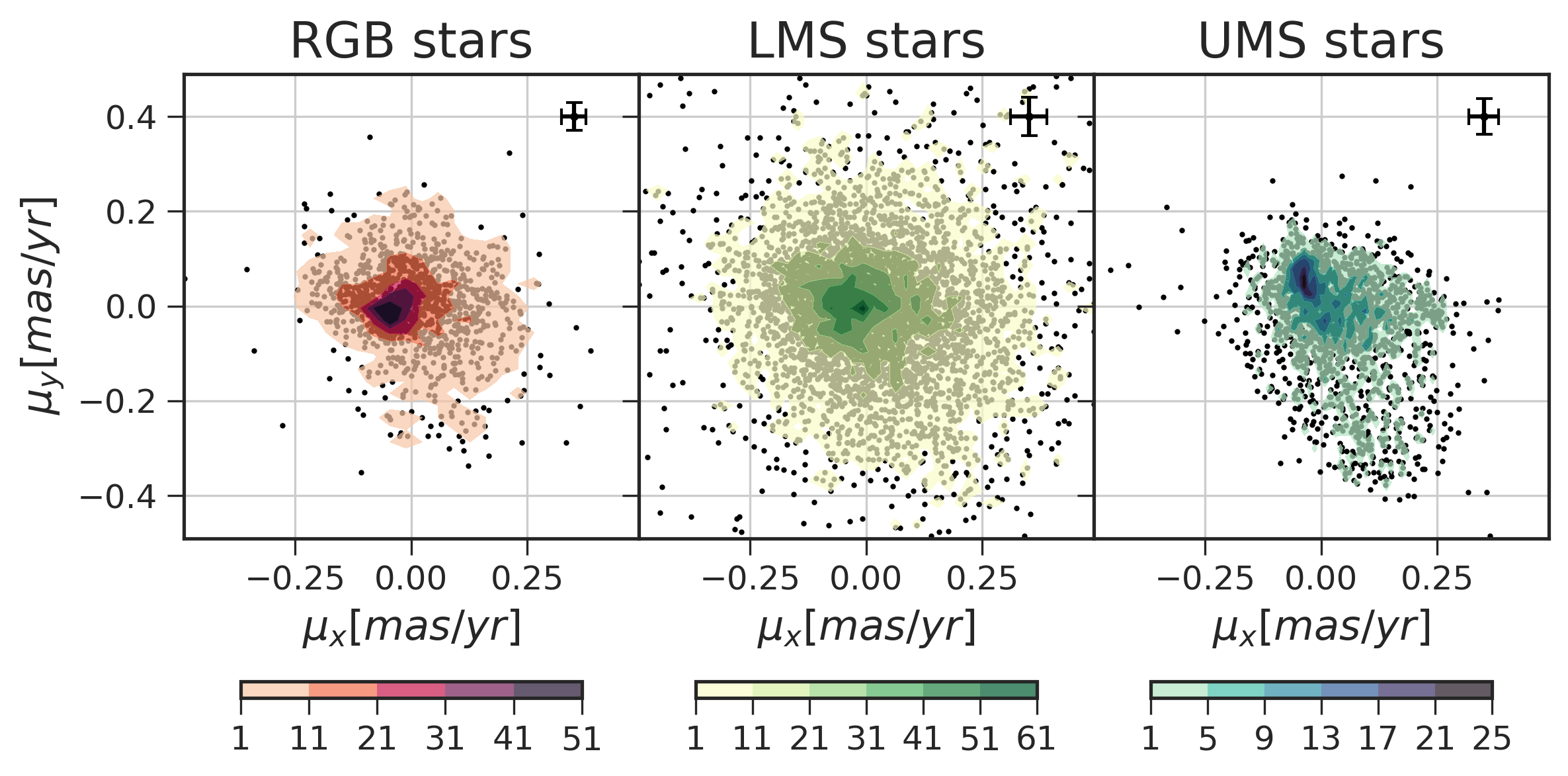}
\caption{PM VPDs measured in the F814W filter for three evolutionary phases, selected using the CMD shown in Fig.~\ref{f:cmd}. RGB stars are shown in the {\it left pabel}. LMS stars are in the {\it central panel} and UMS stars are in the {\it right panel}. For each population, the median error in the $\mu_x$ and $\mu_y$ directions is shown in the upper right corner.}
\label{f:VPD_pop}
\end{figure*}

The inspection of the CMD in Fig.~\ref{f:cmd} reveals a complex history of SF, characterized by episodes of different intensity. A detailed analysis of the region SF history is presented in \citet{Cignoni2011}. Here, we will briefly discuss the properties of the CMD that will be useful for the subsequent analysis.

Above $m_{F814W} \simeq 21$, the CMD shows two well-defined sequences: the blue  ($-0.3\le m_{F555W}-m_{F814W}\le 0.4$) upper main sequence (UMS), and the red giant branch (RGB), with ($m_{F555W}-m_{F814W}\ge 0.8$). The UMS includes intermediate and high-mass stars ($M\gtrsim 3\, M_\odot$) that formed between $\lesssim 2$ and $\sim 600$ Myr ago. The youngest component belongs to NGC~346, while the remaining stars are associated with the SMC field.

The RGB consists of evolved low-mass stars ($\lesssim 2\, M_\odot$) characteristic of a stellar population older than  $\sim 1$ Gyr. The compact group of stars around $m_{F814W}\simeq 18.5$ and  $m_{F555W}-m_{F814W}\simeq 1.0$ is the red clump (RC) and corresponds to the core Helium burning phase for stars in the mass range between $\sim 1$ and $\sim 2.5\,  M_\odot$, with a main-sequence lifetimes ranging between $\sim 1$ and 10 Gyr. The older stellar population includes both SMC and BS90 stars. A few bright ($m_{F814W}<18$) evolved red super-giants connect the UMS to the top of the RC.

\begin{figure*}[htb]
\plotone{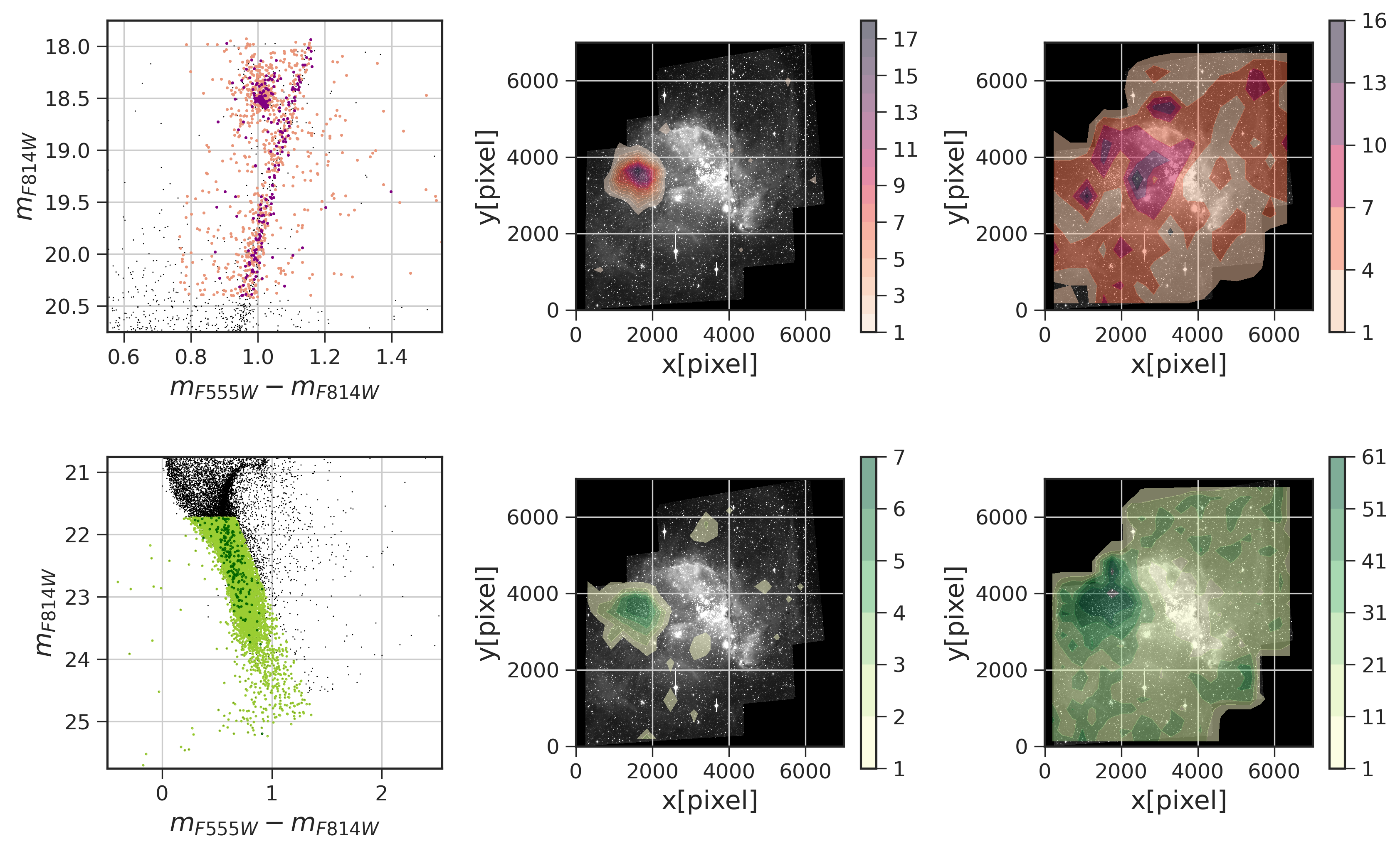}
\caption{Distribution of low-mass stars, as selected from the RGB and LMS VPDs. {\it Upper-Left Panel :} $m_{F555W}-m_{F814W}, m_{F814W}$ CMD centered on the RGB evolutionary phase. Stars that in the RGB VPD are within 0.03 mas/yr from the over-density peak at $\mu_x = -0.04$ mas/yr and $\mu_y = -0.02$ mas/yr are shown in purple, while stars more than 0.08 mas/yr away from the peak are in pink. {\it Upper-Middle Panel :} Spatial distribution of the sources found within 0.03 mas/yr from the over-density peak in the RGB VPD superimposed on the F555W HST image. {\it Upper-Right Panel :} Spatial distribution of the RGB stars found more than 0.08 mas/yr away from the over-density peak in the RGB VPD. {\it Lower-Left Panel :} Portion of the $m_{F555W}-m_{F814W}, m_{F814W}$ CMD centered on the LMS evolutionary phase. Stars that in the LMS VPD are within 0.03 mas/yr from $\mu_x = -0.04$ mas/yr and $\mu_y = -0.02$ mas/yr are shown in dark green, while stars more than 0.08 mas/yr away are in light green. The spatial distribution of the two groups of stars is shown in the {\it Lower-Middle Panel} and {\it Lower-Right Panel} respectively.}
\label{f:rgb_map}
\end{figure*}

The majority of the stars below $m_{F814W}\sim 22$ are low-mass main-sequence (LMS) stars. This sequence includes stars that formed between a few tens of Myr and several Gyr ago and belong either to the SMC field or BS90. NGC~346 stars below $\sim 3\, M_\odot$ are still in the pre-main sequence (PMS) phase, and populate the faint ($m_{F814W}>20$) cloud of red ($m_{F555W}-m_{F814W} > 1.0$) sources to the right of the LMS.

\section{Proper Motions}
\label{s:PMs}

The first step in measuring PM displacements is to define the reference system. We considered basing the absolute astrometric reference frame on the Gaia Early Data Release 3 \citep[EDR3; ][]{Gaia2021} catalog. However, below $m_{\rm F814W} = 18$ (which corresponds to our saturation threshold), there are fewer than 600 stars in EDR3 with position errors $< 1\, {\rm mas}$ and PM errors $< 1\, {\rm mas/yr}$ in common with our catalog. Furthermore, this list includes both the young and old stellar populations, that could have different relative motions. Thus, to create reliable astrometric reference frames, we would have had to project the stars to the positions they had in 2004. Given that the average PM uncertainty for these stars is 0.9 mas/yr, the reference frame of the first epoch would not meet the required astrometric accuracy. We, therefore, decided to follow the same approach used by \citet{Bellini2014} and to create a network of reference sources directly from our catalog. The PM displacement of each source is then measured relative to this reference network.

We measured the PM of NGC~346 stars with respect to the RGB stellar population since the latter can be easily identified in the CMD (Fig.~\ref{f:cmd}) using color and magnitude selections. We used the F814W exposures acquired in the first epoch to determine the average position of the RGB stars and create the reference network. In total, we identified more than 1400 RGB stars with photometric error $<0.1$ magnitudes in both filters and epochs and with position errors $< 1.0\, {\rm mas}$ in both epochs. We then applied the six-parameter global transformations needed to translate stellar positions in each exposure onto the reference frame system. To limit the number of mismatches, we considered only stars whose coordinates in the master-frame match to within 2.5 pixels ($0.125\arcsec$).

To mitigate the effect of small, uncorrected systematics in, e.g., the {\it HST} geometric-distortion solutions and PSF models, we applied local corrections derived from the average residuals between the transformed positions of the closest N=30 RGB reference stars. The median distance of the furthest reference star is 600 pixels ($\sim 30\arcsec$).

\begin{figure*}[htb]
\plotone{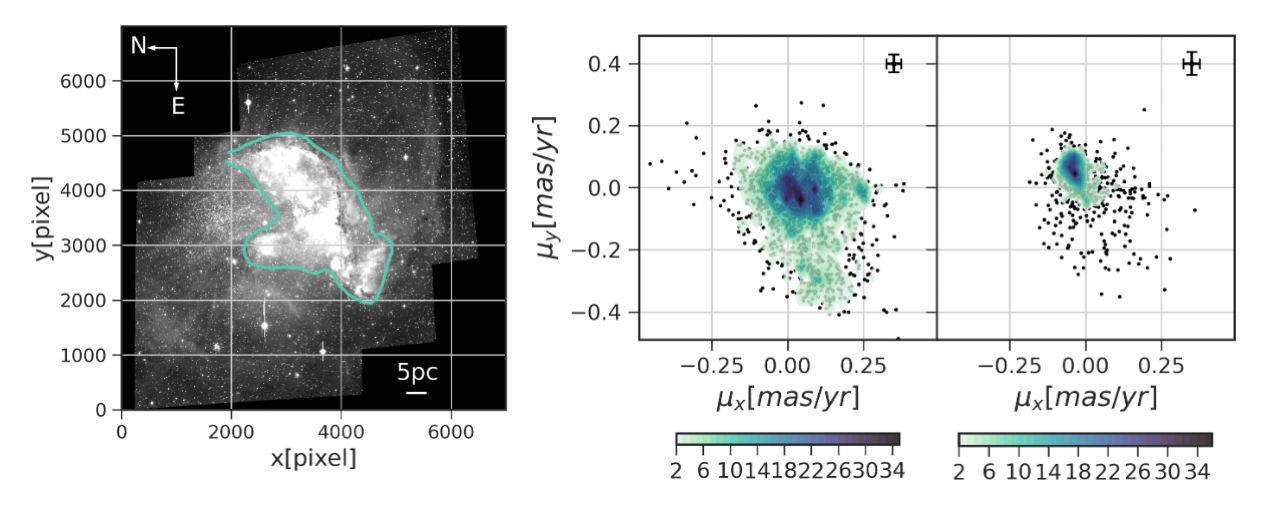}
\caption{VPDs of the PMs for the UMS stars. The {\it left panel} shows the $5.48\times 10^{-17}\, erg/cm^2/\mathrm{\AA}/s^{-1}$ isophotal contour level superimposed on the F555W image. The contour has been used to separate the sources likely associated with the NGC~346--N66 complex from the stars belonging to the young SMC field. The {\it central panel} shows the PM VPD of the young SMC field, and the {\it right panel} shows the VPD of the PMs for the NGC346 stars.}
\label{f:ums_VPDs}
\end{figure*}

For each epoch, we computed the final position of each star as the sigma-clipped mean of the transformed positions of all images in the F814W filter at that epoch. We repeated the same analysis for the data taken in the F555W filter. In each filter, we defined the PM of each star as the difference between its position in the second and the first epoch, divided by the 11 years temporal baseline. We estimated the PM errors by adding the positional errors in each epoch in quadrature and dividing by the temporal baseline.

For the analysis of the PMs, we considered only sources with photometric errors  $<0.1$ magnitude in both filters and epochs and with X and Y positional errors $<1.0$ mas in both epochs. As shown in Fig.~\ref{f:errors}, only sources in the magnitude range $18<m<23$ met both our astrometric and photometric requirements.

Fig.~\ref{f:PMerrors} shows the PM 1-D rms of the PMs as a function of magnitude.
PM measurements in the F555W filter are affected by larger errors because of the smaller number of dithers in the second epoch's southern pointing, and the higher background level, caused by the ionized gas. Therefore, we decided to measure PMs primarily using the F814W filter, although we verified that the F555W observations provide consistent results. After the selections in photometry and position errors, the average PM rms in the F814W filter is $0.06\pm0.02$ mas/yr and $0.07\pm 0.02$ mas/yr in the F555W. For consistency with previous studies of NGC~346 \citep[e.g. ][]{Sabbi2008, Cignoni2011, Gouliermis2014}, in this work we adopted 60.4 kpc as the nominal distance of the system. Thus, the average PM rms\footnote{Depending on the method and studied region, the reported distance of the SMC in the literature varies from less than 40 \citep{Groenewegen2013} to almost 100 \citep{Issa1989} kpc. While it is beyond the scope of this paper to discuss the challenges and merits of the various techniques, we noticed that the uncertainties on the distance of NGC~346 considerably affect the magnitude and the dispersion of all the velocity distributions studied here.} in the F814W filter is $17.2\pm 5.7$ km/s and $20.1\pm 5.7$ km/s in the F555W filter. While the PM uncertainties for most individual stars are too large to resolve their internal motion within the cluster, the averaged PMs for a population of stars or over a spatial subregion of the cluster {\it are} accurate enough to study the internal cluster kinematics, since averaging decreases the uncertainties by $N^{-1/2}$.

\begin{figure}[htb]
\plotone{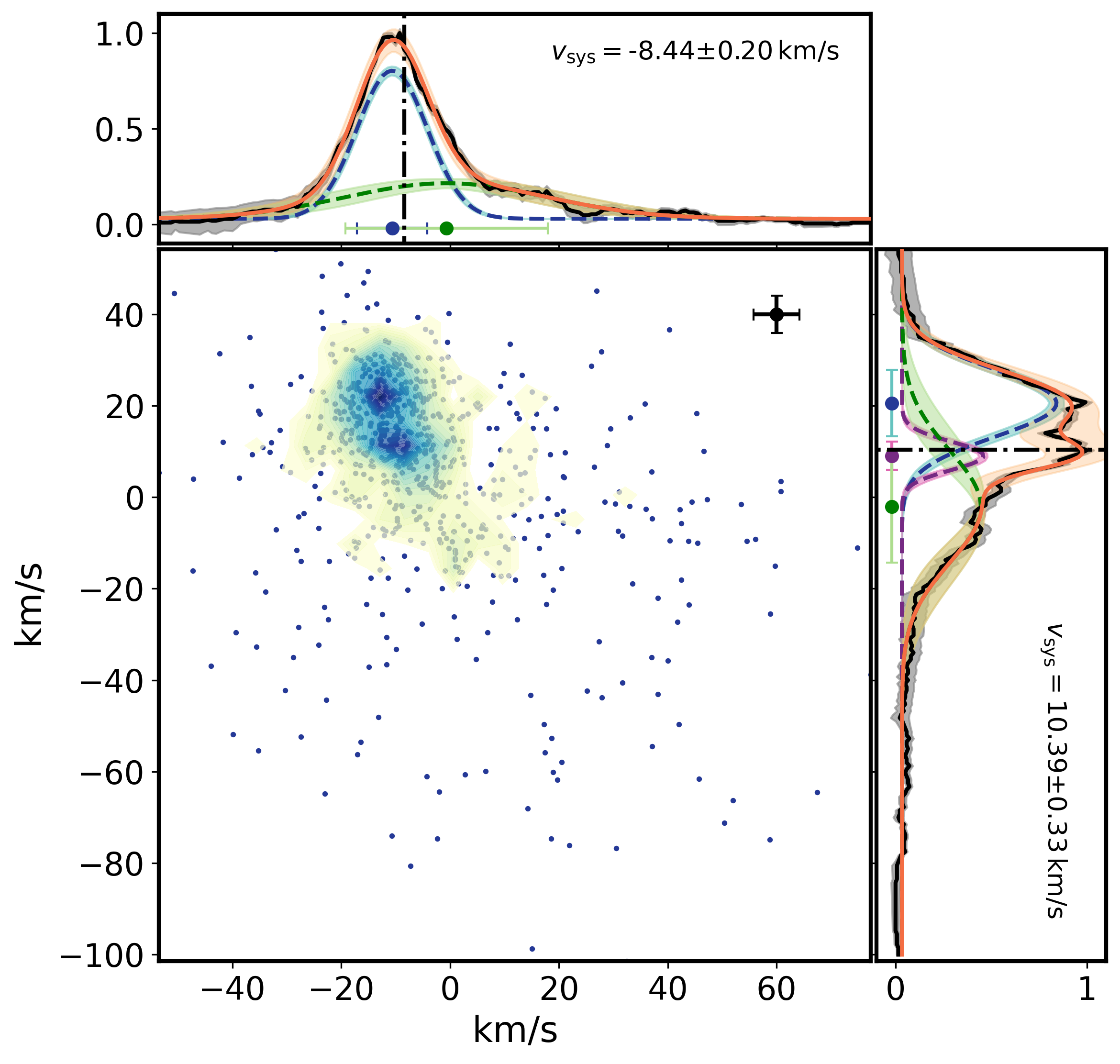}
\caption{{\it Central plot:} VPD of the stars in NGC~346 expressed in km/s instead of mas/yr. The error bars represent the typical standard deviations in a bin. This information was used to determine the size of the binning step.
{\it Upper histogram:} Velocity distribution along the X direction. The systemic velocity (marked by the dashed-dotted black line) is
$v_{sys_x}= -8.44 \pm 0.20$ km/s. The distribution best fit (Orange continuous line) is achieved combining two Gaussians. The center of the stronger component (blue dashed line) is at $v_{x1}=-2.19\pm 0.13$ km/s from NGC~346 systemic velocity, with dispersion $\sigma_{x1}=6.48\pm 0.20$ km/s and number of stars at the peak $a_{x1}=160.14\pm 4.72$. The broader component (green dashed line) is at $v_{x2}=7.80\pm 1.87$ km/s from the systemic velocity, with dispersion $\sigma_{x2}=18.61\pm 2.71$ km/s and $a_{x2}=38.39\pm =4.88$ stars. {\it Side histogram:} Velocity distribution along the Y direction. The systemic velocity (marked by the dashed-dotted black line) is
$v_{sys_y}= 10.39 \pm 0.33$ km/s. The distribution best fit (Orange continuous line) results from the combination of three Gaussian curves. The stronger component (blue dashed curve) is redshifted from NGC~346 systemic velocity by $v_{y1}=10.16\pm 0.38$ km/s with dispersion $\sigma_{y1}=7.30\pm 0.33$ km/s and  $a_{y1}=111.46\pm 5.24$ stars at the peak. The broader component (green dashed line) is at $v_{y2}=-12.47\pm 2.16$ km/s {\bf from NGC~346 systemic velocity}, with dispersion $\sigma_{y2}=12.27\pm 2.29$ km/s and  $a_{y2}=56.78\pm 2.05$ stars. The third component (purple dashed line) is fitted with a Gaussian function with center $v_{y3}=-1.37\pm 0.23$ km/s from NGC~346 systemic velocity, $\sigma_{y3}=3.09\pm 0.30$ km/s, and $a_{y3}=58.90\pm 2.65$ stars.}
\label{f:346_vel}
\end{figure}

The vector-point diagram (VDP, Fig.~\ref{f:VPD_tot}) of the PMs in the NGC~346 region shows a complex structure with indications of multiple peaks in the higher density part of the plots and an elongated structure toward the lower right corner. It is clear from the inspection of Fig.~\ref{f:VPD_tot}, that this plot alone is not sufficient to separate the different populations.

\section{Kinematics of the Stellar Populations}
\label{s:kinem}

The internal dynamics of the SMC is strongly affected by its recent history of interaction with the Large Magellanic Cloud, with the SMC being in the process of tidal disruption \citep{Zivick2018}. The numerous attempts to derive the internal motions of the various SMC components using either line-of-sight \citep{Stanimirovic2004, Harris2006, Evans2008, Dobbie2014} or PM \citep{Kallivayalil2006, Kallivayalil2013, VdM2016, Zivick2018} measurements have thus far provided contradicting results, highlighting how, especially on small scales, younger and older stars in the field of the SMC may have different kinematics.

To better understand the behavior of the various stellar populations in the NGC~346 region, in the following analysis we will take advantage of the fact that for each star we have three primary bits of information:
\begin{itemize}
    \item The position projected on the sky (Fig.~\ref{f:mosaic});
    \item The position in the CMD (Fig.~\ref{f:cmd});
    \item The position in the VPD (Fig.~\ref{f:VPD_tot}).
\end{itemize}
As a starting point, we used the CMD to separate RGB, LMS, and UMS stars, and then analyzed the motion of the three stellar components separately. The VPDs for the three populations are shown in Fig.~\ref{f:VPD_pop}.

\subsection{Low-Mass Stars in the SMC Field}

In this section we will examine the motion of the stars with mass $M\lesssim 2\, M_\odot$ in the SMC field. As discussed in section \ref{stellar pops}, this group of stars can be divided in RGB and LMS stars. The left panel of Fig.~\ref{f:VPD_pop} shows the VDP of the RGB stars. Since the reference frame and the transformations are based on the RGB stars, the weighted average of this population in the VDP is by construction centered on zero {\bf ($\mu_x = -0.0061\pm 0.0003\, {\rm mas/yr}$, $\mu_y= -0.0028\pm 0.0002\, {\rm mas/yr}$)}.

The properties of the RGB stars selected using their positions in the VPD  of the PMs are shown in the three upper plots of Fig~\ref{f:rgb_map}. In the CMD (upper-left panel) sources within 0.03 mas/yr (corresponding to a velocity dispersion of $\sim 8.5$ km/s) from the over-density peak found at at $\mu_x=-0.04\, {\rm mas/yr}$ and $mu_y=-0.02\, {\rm mas/yr}$ are characterized by a small color dispersion. The RC of this populations is compact and well defined, suggesting that this group of stars has similar ages, chemical composition, distance, and it is affected by comparable dust extinction. Spatially, these sources are all concentrated around the nominal position of the 4.5 Gyr old BS90 cluster, as shown by the density contours in the upper-middle panel.

In contrast the stars that are found more than 0.08 mas/yr away from the density peak are likely part of the old field of the SMC. In this case the RC shows a broader dispersion in magnitude, suggesting that the population could be distributed along a broader line of site, and include stars that formed over a larger period of time. This is further supported by the fact that the brighter portion of the RC terminates in the V-like shape formed by the red and blue edges of the helium-burning phase typically observed in the CMDs of stellar populations younger than 1 Gyr, suggesting that this component formed stars for several billion years. These sources cover the entire studied region (as shown in the upper right panel). The impact of BS90 on the PM measurements is negligible: in fact after removing the RGB stars found within $75\arcsec$ from the center of BS90, the weighted average of the RGB in the VPD remains $\mu_x=0.0020\pm 0.0001\, {\rm mas/yr},\, \mu_y= 0.0002\pm  0.0002\, {\rm mas/yr}$.

\begin{figure*}[htb]
\plotone{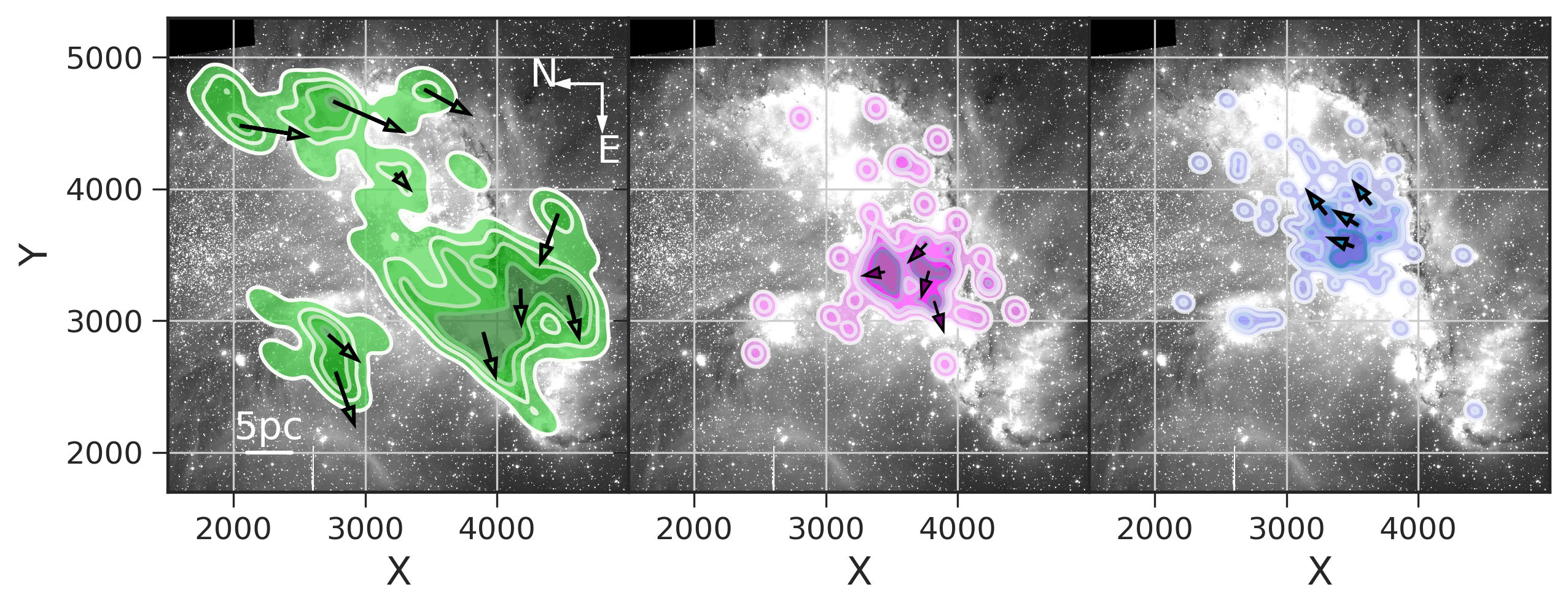}
\caption{Spatial distribution of the stars found within $0.5\, \sigma$ from the three Gaussian peaks found in the {\it side histogram} of Fig.~\ref{f:346_vel}. The {\it left panel} shows the distribution of the stars associated with the broader green Gaussian, the {\it central panel} the distribution of the stars associated with the narrow magenta Gaussian, and the {\it right panel} the distribution of the stars found under the blue Gaussian. The arrows indicate the average motion of the stars  found in the higher density regions, after the mean motion of NGC~346 has been subtracted.}
\label{f:directions}
\end{figure*}

The middle panel of Fig.~\ref{f:VPD_pop} shows the PM VPD of the LMS stars. In addition to stars from the old field and BS90, this plot also includes the SMC's younger ($\sim 50-100$ Myr to $\lesssim 1$ Gyr) component. The VPD is centered at $\mu_x,  \mu_y = 0,0$ mas/yr, but the presence of BS90 can still be recognized in the second higher density contour. The properties of the LMS stars are highlighted in the three lower panels of Fig~\ref{f:rgb_map}. As for the RGB stars, the sources found within  0.03 mas/yr from  $\mu_x=-0.04$, $\mu_y=-0.02$ are concentrated around the center of BS90 and in the CMD are distributed along a tight and well-defined sequence, suggesting that they all belong to a coeval stellar population, with negligible distance variation. On the contrary, the remaining stars show a uniform spatial distribution and a much broader LMS, as expected for field stars, affected by different reddening, distance, and possibly metallicity. These results imply that BS90 has a velocity of $v_x=-11.5\pm 0.5,\, v_y=-5.7\pm 0.7$ km/s (corresponding to $v_{RA}=5.34\pm 0.70,\, v_{Dec}=11.67\pm 0.50$ km/s) relative to the more broadly distributed field RGB population.

\begin{figure*}[htb]
\plotone{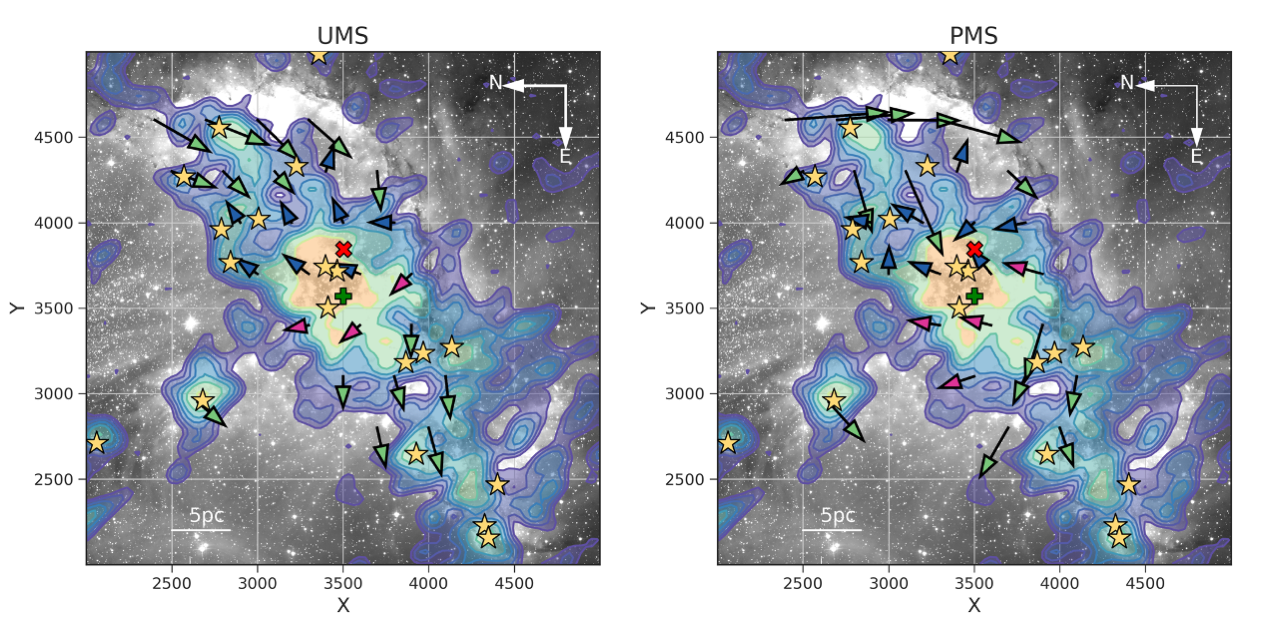}
\caption{Motion of the stars within the isophotal contour shown in Fig~\ref{f:ums_VPDs}. The {\it left panel} shows the motion of the UMS stars, while the {\it right panel} is for the PMS stars. Each arrow indicates the average direction of the stars found within a $200\times 200$ pixel grid, after subtracting the total NGC~346 motion. The arrow length, increased 10,000 times, corresponds to the measured displacement. In the left panel the contamination of the SMC young field was statistically removed. Green, blue and magenta colors were used to highlight different directions. Yellow star symbols mark the position of the YSOs identified by \citet{Sewilo2013}. The green + symbol highlights the position of the photometric center, while the rotation center is marked with the red X. The density contours highlight the spatial distribution of the PMS stars.}
\label{f:motions}
\end{figure*}


\subsection{Intermediate and High-Mass Stars}

The distribution of the UMS stars in the VPD (Fig.~\ref{f:VPD_pop}, right panel) is quite different from those observed for the RGB and LMS stars. This is the only diagram that shows an elongated structure in the upper left part ($\mu_x\simeq -0.025\, {\rm mas/yr}$, $\mu_y\simeq 0.03\, {\rm mas/yr}$), associated with the NGC~346--N66 system.

To better separate NGC~346 stellar content from the young SMC field,  we used the isophotal contour level $5.48\times 10^{-17}\, erg/cm^2/\mathrm{\AA}/s$ (Fig.~\ref{f:ums_VPDs} - left panel). The central panel shows the VPD of the stars outside the isophotal level. The majority of these stars most likely belong to the SMC field. The right panel shows the VPD of the stars inside the isophotal contour. The plot here shows two separate peaks, embedded in an extended halo. The majority of these sources likely belong to NGC~346.

The VPD of the SMC young field is clearly shifted to the right of NGC~346 and peaks at $\mu_x=0.043\, {\rm mas/yr}$, $\mu_y=-0.040\, {\rm mas/yr}$, with an extended and broad tail towards the lower-right part of the diagram. Contrary to what is found by \citet{Zivick2018}, both the Kolmogorov-Smirnov test and the Cram\'er-von Mises criterion reject the hypothesis that young (UMS) and old (RGB) field stars have similar kinematics. The discrepancy between Zivic's and our results is likely due to the recent star formation history of the NGC~346 region. \citet{Cignoni2011} reported an elevated star formation rate over the past 100 Myr compared to the rest of the SMC. We suggest that the elevated and localized activity occurred in the past $\sim 100$ Myr likely dominates the kinematics of the young field stars, and is the culprit for the observed difference.

\begin{figure}[htb]
\plotone{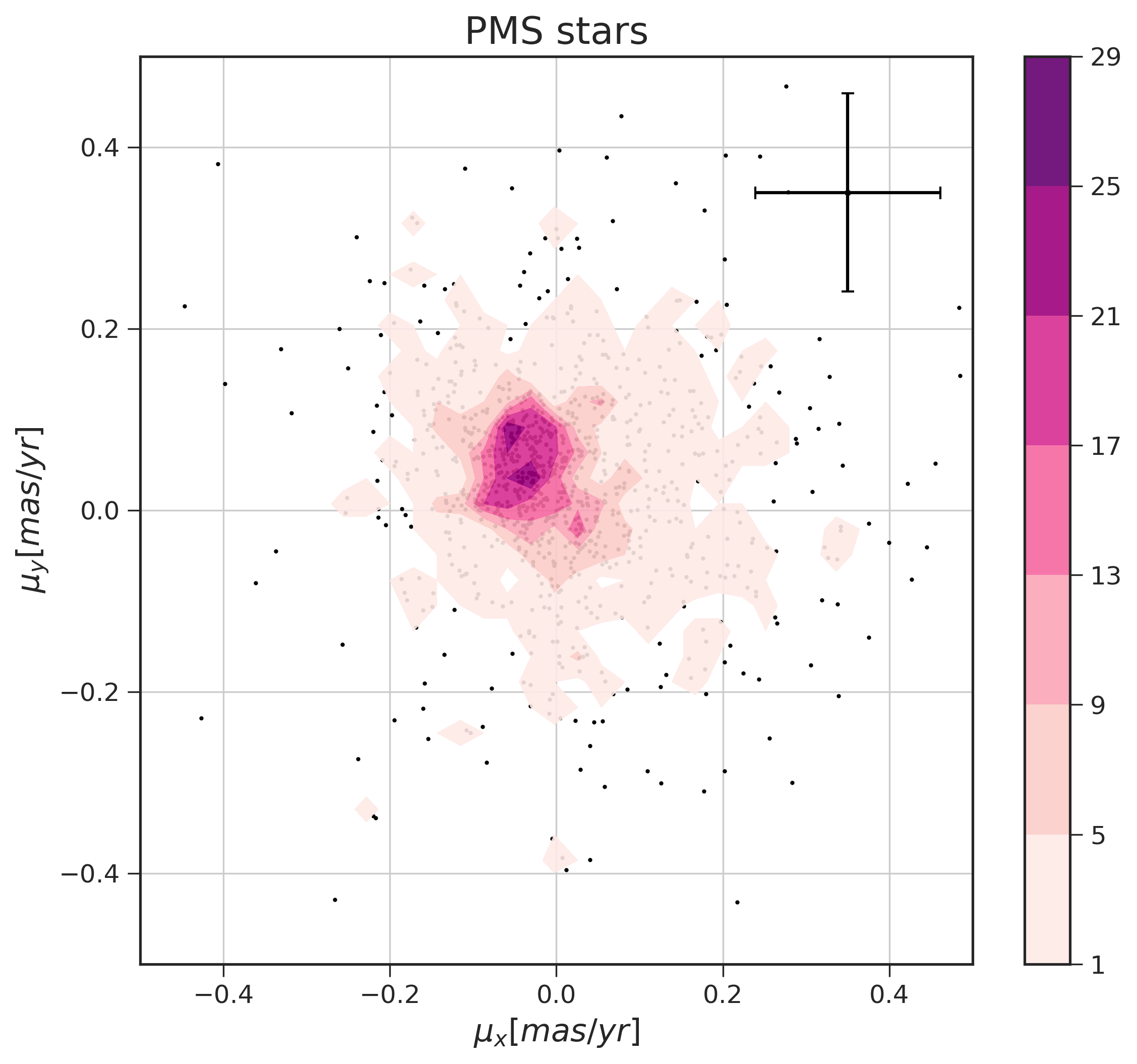}
\caption{PM VPD of the PMS stars, measured in the F814W filter. The median error in the $\mu_x$ and $\mu_y$ directions is shown in the upper right corner}
\label{f:PMS_PMs}
\end{figure}

\section{NGC~346}
\label{s:n346}

\subsection{Upper Main Sequence Stars}

\begin{figure*}[htb]
\plotone{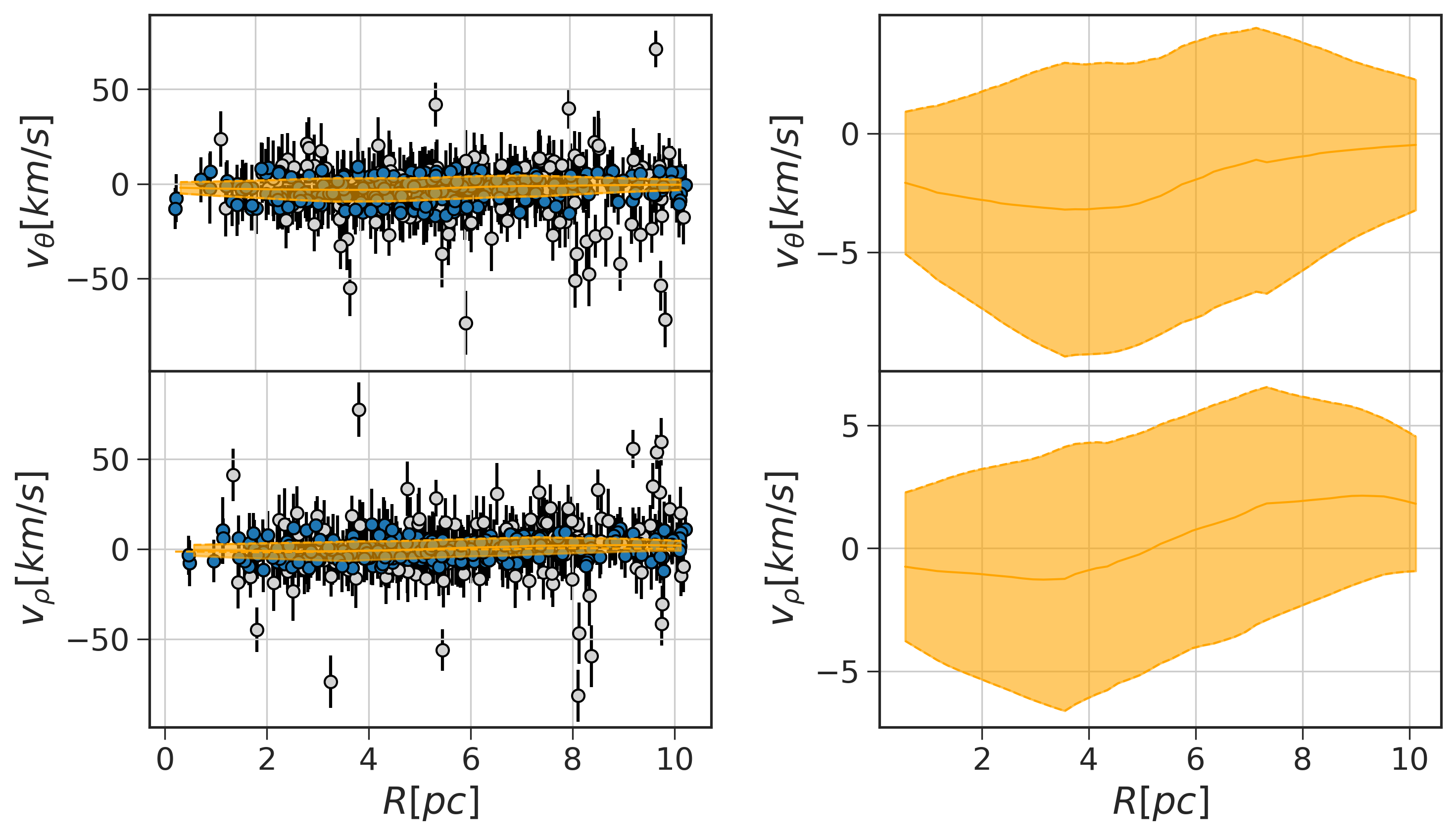}
\caption{{\it Left panels:} $v_{\theta}$ ({\it upper panel}) and $v_{\rho}$ ({\it lower panel}) components of the velocity as a function of the distance from the rotation center. Sources that lie within $2\, \sigma$ from ($v_{\theta},v_{\rho}=0,0$) are shown in blue, and the remaining sources are in grey. The continuous yellow line marks the mean value of the distribution. The dashed yellow lines mark the + and -1 standard deviation. {\it Right Panels:} Zoom in of the $v_\theta$ and $v_\rho$ $1 \sigma$ variation as a function of the distance from the rotation center.}
\label{f:vr_vt}
\end{figure*}


Fig.~\ref{f:346_vel} shows the velocity distribution of the NGC~346 stars in km/s. The systemic velocity of NGC~346 with respect of the RGB stars (defined as the median of the velocities of all the sources within the isophotal contour shown in the left panel of Fig.~\ref{f:ums_VPDs}) is $v_{sys,x}=-8.44 \pm 0.20\, {\rm km/s}$ and $v_{sys,y}=10.39 \pm 0.33\, {\rm km/s}$ (which correspond to $v_{sys,RA}=-10.64 \pm 0.33\, {\rm km/s}$ and $v_{sys,Dec}=8.12 \pm 0.20\, {\rm km/s}$).

The distribution of the PMs appears double-peaked and surrounded by an extended halo. To confirm that the two peaks are not an artifact of a too aggressive binning and take into proper account measurement errors, we ran an extensive Markov chain Monte Carlo (MCMC) Bayesian inference comparing a single Gaussian with the sum of two and three functions. Both the Akaike \citep[AIC; ][]{Akaike1974}, and the Bayesian information criterion \citep[BIC; ][]{Schwarz1978} favor three Gaussians (two narrow peaks + a broader component) in the Y-direction. In the X-direction, the two narrow peaks are almost aligned; thus, we tested only the one and two peak hypotheses. The favored results are shown in the two histograms in Fig.~\ref{f:346_vel}.

Although the test run above is not a formal fitting of the distribution of motions, the parameters of that define the Gaussian functions favored by the Bayesian inference can be used to select the stars that likely belong to the different moving groups. In particular, Fig.~\ref{f:directions} shows the spatial distribution of the stars found within $0.5\, \sigma$ from the Fig.~\ref{f:346_vel} Gaussian curve peaks. The component characterized by the broader velocity dispersion ($\sigma_{x2}=18.61 \pm 2.71,\, \sigma_{y2}=12.27\pm 2.29$ km/s) is also the most extended and covers the entire NGC~346 complex (left panel). With the respect to the NGC~346 systemic velocity, these stars are moving from northwest to southeast with a mean velocity $v_{x2}=7.80 \pm 1.87, \, v_{y2}=-12.47\pm 2.16$ km/s.

The stars associated with the magenta (central panel) and the blue (right panel) Gaussian curves are clustered within the inner  8.8 pc from the photometric center of NGC~346.  With respect to the NGC~346 median motion, most of the stars in the magenta component seem to move along an arch from  South to North. After the subtraction of the NGC~346 systemic velocity, the mean velocity in the Y-direction is  $v_{y3}=-1.37\pm 0.23$ km/s, and the velocity dispersion $\sigma_{y3}=3.09\pm 0.30$ km/s. The blue component moves in the opposite direction to the green one, from the center to the northwest. The mean velocity in the Y-direction is redshifted compared to NGC~346 median velocity by $v_{y1}=10.16\pm 0.38$ km/s, with velocity dispersion $\sigma_{y1}=7.30\pm 0.33$ km/s. Combined, the two components along the X-direction are on average blueshifted compared to NGC~346 systemic velocity by $v_{x1}=-2.19\pm 0.13$ km/s and $\sigma_{x1}=6.48\pm 0.20$ km/s.

Fig.~\ref{f:directions} reveals a complicated internal kinematics for NGC~346. The system appears dynamically hot, with large velocity dispersion and sudden changes in directions.

The comparison between the VPDs of the SMC young field and NGC~346 UMS stars (Fig.~\ref{f:ums_VPDs}) shows a certain level of overlap between the motions of the two populations. To address the impact of field contamination, we measured the stellar density of UMS stars between $17.9<m_{F814W}<20.4$, that meet our photometry and astrometry requirements, at a distance greater than $R>150\arcsec$ (corresponding to $R\simeq 44\, {\rm pc}$) from the center of NGC~346. This yield a stellar density of $<0.01$ star per squared arcsecond. We then divided the region inside the isophotal contour shown in Fig.~\ref{f:ums_VPDs} in a $200 \times 200$ pixel grid, and we measured the density of the UMS stars in each cell. Assuming a uniform distribution and 100\% completeness for the UMS stars of the SMC young field, we estimate that the field contamination in the various cells ranges from 5 to 30\%, with an average value of 12\%. We noticed that this is likely an upper limit, since the NGC~346 region is likely more incomplete, due to the higher crowding and background level, than the outer parts of the image.

We statistically removed the contamination of the SMC young field from NGC~346 systemic velocity with the respect of the RGB, which became $v_{sys,x}=-9.87\pm 0.23\, {\rm km/s}$ and $v_{sys,y}=11.82\pm 0.37\, {\rm km/s}$, and in equatorial coordinates it corresponds to $v_{sys,RA}=-11.64\pm 0.37\, {\rm km/s}$ and $v_{sys,y}=10.07\pm 0.23\, {\rm km/s}$. As shown in Fig.~\ref{f:motions} (left panel), after statistically removing the field contamination, the center, and possibly the entire upper part, of the NGC~346 complex, seems to rotate clockwise, once the systemic velocity of NGC~346 is subtracted. The lower half of the system appears to move away from the complex as if NGC~346 is breaking into two parts.

\subsection{Pre Main Sequence Stars}

Because of NGC~346 young age, stars below $\sim 3\, M_\odot$ are still in the PMS evolutionary phase. With the exception of a few small clumps, these stars are mainly concentrated within the isophotal contour shown in Fig.~\ref{f:ums_VPDs}.

Most of the PMS stars do not meet our PM selection criteria, likely because of the elevated level of stellar crowding and high background. However, in Fig.~\ref{f:cmd} CMD, PMS stars fainter than $m_{F814W} > 21.2$ represent the cleanest sample of NGC~346 stars. We therefore decided to relax the selection criteria (blue stars in Figs.~\ref{f:errors} and \ref{f:PMerrors}) to perform a second, independent, analysis, although affected by larger uncertainties, of NGC~346 internal kinematics.

Fig.~\ref{f:PMS_PMs} shows the VPD for the PMS stars with photometric rms $< 0.2$ in both the F555W and F814W filters, and with positional rms in F814W $< 3$
mas in both epochs (purple sources in the CMD shown in Fig.~\ref{f:cmd}). As in the case of NGC~346 UMS stars, the plot shows two clear peaks, surrounded by an extended halo, in the upper left part of the diagram. A third, weaker peak is visible towards the lower right part of the plot.

As we did for NGC~346 UMS stars, we divided the PMS stars in a $200 \times 200$ pixel grid and for each cell we measured the average motion of the PMS stars, corrected for the cluster's systemic velocity. Since this group of stars was selected using the CMD, where the field contamination is negligible, no further correction was needed. Although the plot appears noisier than in the case of the UMS stars, the two distributions are qualitatively very similar. Both diagrams suggest that the upper part of the NGC~346 is rotating. The southern end of the system, on the other end, displays an entirely different motion, resembling an outflow.

\subsection{Velocity and Velocity Dispersion Profiles}
\label{s:polar}

We derived for each UMS star in NGC~346 the radial ($v_\rho$) and tangential ($v_\theta$) velocity components to study the profile of the cluster mean velocity and velocity dispersion. To start, we consider as the rotation center NGC~346's photometric center  ($X=3501.43,\, Y=3568.69$, corresponding to $R.A.=14.7706984,\, Dec=-72.1775412,\,ICRS$), whose position in Fig.~\ref{f:motions} is marked by the green cross symbol.

We subtracted the system's mean PM from each star, and then we derived the $v_{\rho}$ and $v_{\theta}$ components for each source. In this reference system, within the inner $R\sim 11-13$ pc, both $v_\rho$ and $v_\theta \propto r$, with fitted slope $\sim 0.3$ and $\sim 0.2$ km/s/pc, respectively, suggesting that the star-forming regions is rotating as a rigid body, and expanding.

Given the evidence for rotation, we then proceeded to identify the center of rotation, which need not coincide with the photometric center. We looped through a $20\times 20$ pixel grid, and for each grid element, after having subtracted the mean PM of the stars found within 10 pc, we derived the sigma-clipped mean $\overline{v_\theta}$ value of the stars found within the same radius. We considered as center of rotation the grid point that gave the largest $|\overline{v_\theta}|$, found at $RA=14.7575892,\, Dec=-72.1779791,\, ICRS$ ($X=3502.35,\, Y=3845.09$), $\sim 4.5$ pc west of the photometric center. In Fig.~\ref{f:motions} the position of the rotation center is marked with a red X symbol.

Fig.~\ref{f:vr_vt} shows $v_\theta$ and $v_\rho$ as a function of distance from NGC~346's rotation center. Despite the large dispersion, $v_\theta$ (upper-right panel) appears to increase between $\sim 0.5$ and 3.5 pc (where it reaches the maximum value $v_{\theta_{max}}= -3.2\, {\rm km/s}$) at $0.2\, {\rm km/s/pc}$, and then progressively decrease to almost zero at 10 pc ($v_{\theta_{10\, pc}}=-0.5\, {\rm km/s}$).

Within the inner 4 pc $v_\rho$ is negative and almost constant, indicative of an inflow in the inner regions. Instead, between $\sim 3.7$ and $\sim 8$ pc it increases with the distance with fitted slope $\sim 0.2$ km/s/pc, indicative of expansion of the outer regions (Fig.~\ref{f:vr_vt}, lower-right panel). Similar expansions have been observed also in some young Milky Way star clusters \citep{Kuhn2019} and predicted by numerical simulations of the stellar feedback impact on young star cluster evolution \citep{Grudic2022}.

The combination of rotation and inflow in the central regions is suggestive of inspiraling motion. Fig.~\ref{f:vr_vt_map}
shows the spatial distribution and direction of motion of NGC~346's individual UMS stars in the new reference system. In this plot the majority of the stars above $Y=-500$ pixels is spiraling from the north east towards the rotation center. A logarithmic spiral is overdrawn to guide the eye. The distribution of YSOs and sub-clusters aligns well with this motion.

We speculate that in proximity of the center, the stream of spiraling stars bend behind NGC~346 and re-emerge below $Y=-500$ as one of the two tails of stars that are moving toward the southeast, away from NGC~346. The dashed red line in Fig.~\ref{f:vr_vt_map} suggests the possible path followed by these sources.

A narrow stream of stars (indicated by the dashed blue line) and YSOs seems to connect the bright small cluster to the east of NGC~346 \citep[Sc-13 in the classification by ][]{Sabbi2007} to the outer tail. These streams, inflows, and outflows of stars are reminiscent of the filament-like structures that are feeding the growth of star-forming regions in the hierarchical collapse models.

Hydrodynamic simulations of turbulent molecular clouds predict that rotation is a common characteristic of embedded massive \citep[$> 1000\, M_\odot$,][]{LeeHennebelle2016, Mapelli2017,Ballone2020} star clusters, with the rotation curves increasing from the center, peaking approximately at 1-2 half mass radii, and then decreasing in the outer regions, where they can become slightly retrograde \citep[e.g.][]{Tiongco2017}. Over time, the rotation of star cluster is expected to decrease because of stellar mass loss and two-body  relaxation \citep{Einsel1999, Ernst2007, Kim2008, Hurley2012, Tiongco2017}.

Evidence for internal rotation in NGC~346 has been also found over the central $\sim 1\, arcmin^2$ by Zeidler et al. (2022 - submitted) using line-of-sight velocities obtained with the ESO Multi-Object Spectroscopic Explorer \citep[MUSE,][]{Bacon2010}. Similarly to what we observed in the PM distribution, the MUSE dataset shows two statistically significant velocity groups, and a possible third one. Like the magenta and blue groups shown in Fig.~\ref{f:directions}, one velocity group is more centrally concentrated, while the other one appears elongated along the northwest-southeast direction. As in the case of the PMs, the more centrally concentrated group is rotating with an angular velocity of $\Omega = -0.41 \pm 0.07\,{\rm Myr}^{-1}$, which translates to $v_{\rm rot,RV}^{5\,{\rm pc}} = -1.98\pm0.34\,{\rm km}/{\rm s}$ and $v_{\rm rot,RV}^{10\,{\rm pc}} = -3.95\pm0.67\,{\rm km}/{\rm s}$, at a radial distance from the center of 5 and 10\,pc, respectively, comparable to the results of this work.
Unfortunately the MUSE field of view is too small to provide any information about the motion of the NGC~346 outskirts.

\section{Discussion \& Conclusions}
\label{s:Concl}

The star-forming complex NGC~346 presents several challenges to cluster formation models. The sound crossing time is $t_S = 50\, pc/c_S \sim 2\times 10^8$ Myr; thus, the two edges of the NGC~346 parental cloud were not in thermal contact with each other on a freefall collapse timescale and should therefore present considerable age differences. However, all NGC~346 stars fall in the age range 2-6 Myr \citep{Evans2006, Sabbi2007, Cignoni2013, Dufton2019}, raising the question of how star formation developed in such a synchronized manner over such a wide scale. The Upper Scorpius OB association  \citep{Preibisch1999}, the Serpens molecular clouds \citep{Herczeg2019}, and the NGC~7000 and IC~5070 Nebulae \citep{Kuhn2020} pose similar challenges to all the models of cloud collapse that do not include an accelerating star formation rate.

The spatial and kinematic clustering in young stellar systems can provide valuable constraints on the conditions that led to the onset of star formation \citep[e.g.,][]{Elmegreen2002, Parker2014, Kuhn2019}. Therefore, we used two epochs of deep {\it HST/ACS} observations with a temporal baseline of 11 years {\bf (\dataset[10.17909/y1cf-m826]{https://doi.org/10.17909/y1cf-m826})} to measure the PM displacements of the stars in the star-forming region NGC~346, and reconstruct the cluster kinematics.

\begin{figure}[htb]
\plotone{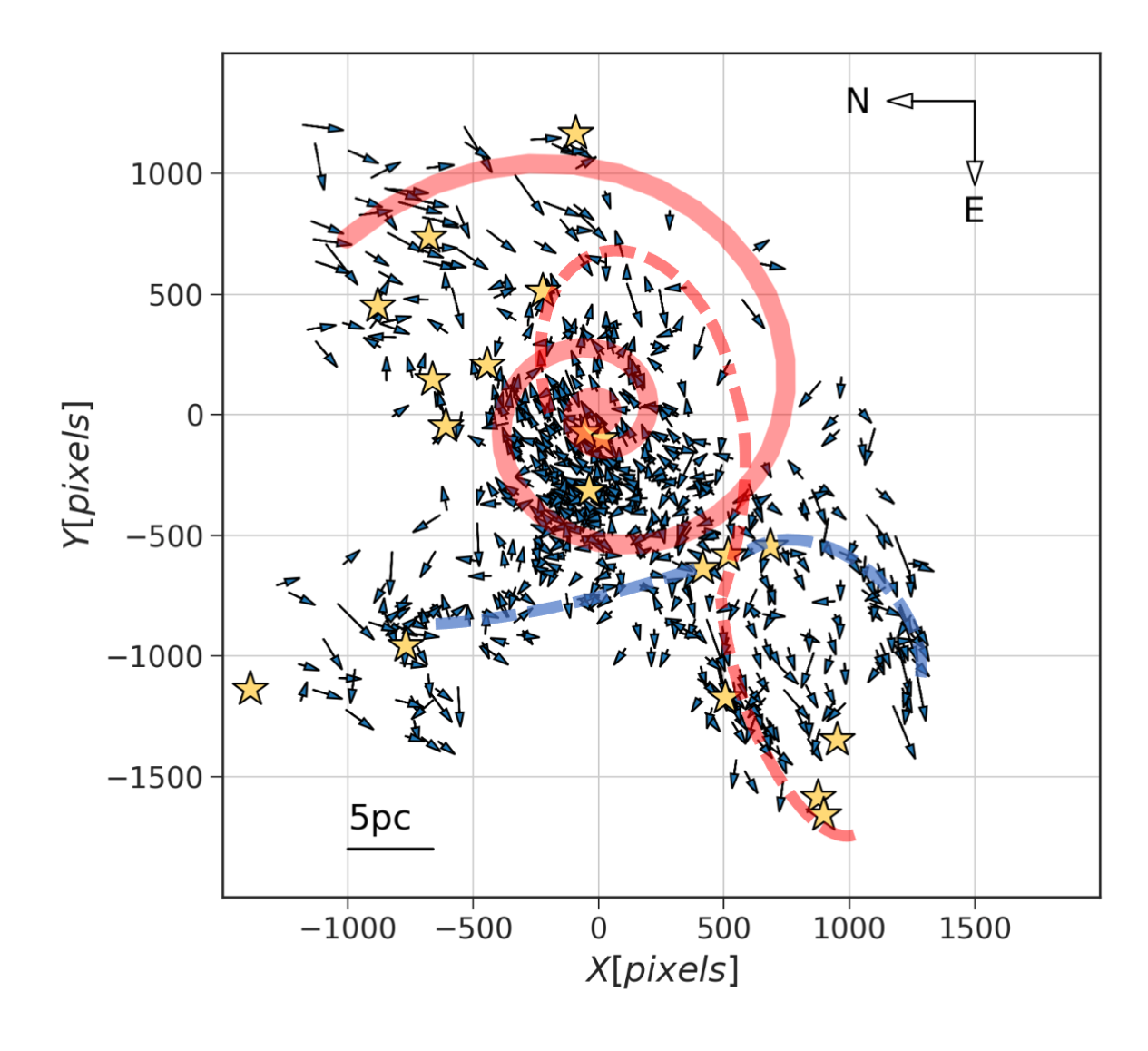}
\caption{Spatial distribution of NGC~346 UMS stars. The arrows indicates the direction of the motion with the respect of the rotation center, while their length is proportional to their velocity. As in Fig.~\ref{f:directions} the yellow stars mark the position of the massive YSOs. The red logarithmic spiral, described by the equation $x = cos(\theta)e^{0.21\theta}$
$y = sin(\theta)exp^{0.21\theta}$, is not a fit, and has been superimposed only to guide the eye. The dashed line connect the stars in the center to the tail of stars that appear to escape NGC~346 from the south west. The dashed blue line highlight the stream of stars that connect the small cluster to the west to the southern tail. }
\label{f:vr_vt_map}
\end{figure}

\begin{itemize}
\item Our analysis of NGC~346 stellar PMs reveals a complex pattern of inflows and outflows. The star-forming region's inner $\sim 10$ pc is rapidly rotating, following a spiraling movement towards the center from the outer north-east end. Increasing rotation with the distance from the center is commonly found in the most up-to-date hydrodynamic simulations of turbulent molecular clouds \citep{LeeHennebelle2016, Mapelli2017, Ballone2020}.

\item NGC~346 is characterized by a clumpy structure, with at least 15 different sub-clusters and asterisms \citep{Sabbi2007,Hennekemper2008}, that often host recently formed YSOs \citep{Simon2007, Sewilo2013}. When we compared the position of sub-clusters and massive YSOs with NGC~346 motion patter, we found that over-densities tend to concentrate at the interface of significant changes in the coherence of the motion field, where we expect to see substantial gas friction and compression. This behaviour appears to be in good agreement with the predictions of rapid star formation in turbulent molecular cloud scenarios \citep[e.g.,][]{Klessen2000, Bonnell2002, Bonnell2003}.

\item NGC~346 shows a large velocity dispersion, and it is increasingly expanding with the distance from the center. Similar behavior is an expected outcome, for example, of global hierarchical collapse models \citep{Vazquez2017, Vazquez2019},  where the collapse of the parental cloud starts as small-scale events within a larger structure and culminates after a few Myr as a series of filamentary flows that accrete onto massive central clumps. These ``river-like'' structures drive the system accretion motions ``down the gravitational potential'' and are responsible for the late appearance of massive stars and their segregation in the central clumps. In agreement with this picture, the PMs studied in this paper revealed longitudinal inflows and outflows that extend over tens of parsecs. These motions could be at the origin of NGC~346's elevated degree of mass segregation reported by \citet{Sabbi2008}.

\item The inspection of the motion field of both UMS and PMS stars indicates that the upper part of NGC~346 is rotating along a coherent spiral-like pattern (Fig.~\ref{f:vr_vt_map}). Both \citet{Cignoni2011} and \citet{Dufton2019} noted that the massive stars in the center of the system likely formed a couple of million years later than the outer part of the cluster. We speculate that the spiraling motion has been feeding this more recent episode of SF.

\item  On a broader scale, our PMs analysis also showed a difference in the relative motions of young ($< 600\, {\rm Myr}$) UMS and old ($>1\, {\rm Gyr}$) RGB field stars. From the analysis of line-of-sight velocities of stars and gas \citet{Stanimirovic2004, Harris2006, Evans2008, Dobbie2014} proposed for the SMC possible age-dependent internal kinematics. On the other hand, \citet{Zivick2018} did not find significant differences between the PMs of UMS and RGB stars in 30 HST fields. We believe that the elevated SF rate in the NGC~346 region over the past $\sim 100\, {\rm Myr}$ is the main reason behind the difference between our and Zivick's findings. In particular, our results show how on a small scale, the specific recent SF history of a region likely dominates the local kinematics. This test highlights the need for a uniform-coverage high-precision astrometric study of the entire SMC to reconstruct the galaxy's internal dynamic and evolution through its history of interactions with the LMC and the MW.
\end{itemize}

Our new measurements for the NGC~346 region represent the first attempt to infer the conditions that induced star formation in a metal-poor giant molecular cloud. We demonstrated that the velocity field of the star-forming region is complex and well matches the conditions predicted by recent hydrodynamic simulations and hierarchical collapse models.
The similarities between our results and those found for star-forming regions in the Milky Way suggest that the differences in the cooling conditions due to the different amounts of metallicity and dust density between the SMC and our Galaxy are too small to alter significantly the process of star clusters assembly and growth. By extending this type of study to a broader range of star formation rates and stellar densities and by comparing the results to theoretical predictions, we can gain new insights into star formation and open a new pathway toward validating or repudiating specific models.

\begin{acknowledgments}

We thank an anonymous referee for carefully reading this manuscript and helping us improve the quality of the paper. We are grateful to M. Gieles, L. Strolger, and R. Klessen for the useful suggestions and discussions.
The HST observations used in this paper are associated with program Nos. 10248 and 13680. The specific observations analyzed in this paper can be accessed from MAST at the Space Telescope Science Institute via \dataset[10.17909/y1cf-m826]{https://doi.org/10.17909/y1cf-m826}. Support for the program 13680 was provided through a grant from the Space Telescope Science Institute. This work is based on observations obtained with the NASA/ESA Hubble Space Telescope, at the Space Telescope Science Institute, which is operated by the Association of Universities for Research in Astronomy, Inc., under NASA contract NAS 5-26555. This work has made use of data from the European Space Agency (ESA) mission {\it Gaia} (\url{https://www.cosmos.esa.int/gaia}), processed by the {\it Gaia} Data Processing and Analysis Consortium (DPAC, \url{https://www.cosmos.esa.int/web/gaia/dpac/consortium}). Funding for the DPAC has been provided by national institutions, in particular the institutions participating in the {\it Gaia} Multilateral Agreement.

\end{acknowledgments}

%

\vspace{5mm}
\facilities{HST(ACS)}


\software{astropy \citep{astropy:2013, astropy:2018}
           }

\bibliography{NGC346}{}
\bibliographystyle{aasjournal}



\end{document}